\documentstyle[aps,prl,twocolumn,floats,epsf]{revtex}
\begin{document}
\draft
\title{Absence of hysteresis at the Mott-Hubbard metal-insulator transition \\
in infinite dimensions}
\author{J. Schlipf$^{\, (1)}\!$, M. Jarrell$^{(2)}\!$, P. G. J. van
Dongen$^{(1)}\!$, N. Bl\"{u}mer$^{\, (1)}\!$, S. Kehrein$^{(3)}\!$, 
Th.\ Pruschke$^{(4)}\!$, and D. Vollhardt$^{(1)}$}
\address{$^{(1)}$Theoretische Physik III, Universit\"{a}t Augsburg,
         86135 Augsburg, Germany;
         $^{(2)}$Department of Physics, University of Cincinnati, Cincinnati, 
         Ohio 45221;
         $^{(3)}$Physics Department, Harvard University, Cambridge, MA 02138;
         $^{(4)}$Institut f\"{u}r Theoretische Physik, Universit\"{a}t
         Regensburg, 93040 Regensburg, Germany\\ {\rm(February 19, 1999)}}

\address{~
\parbox{15cm}{\rm
\medskip
The nature of the Mott-Hubbard metal-insulator transition in the
infinite-dimensional Hubbard model is investigated by Quantum Monte Carlo 
simulations down to temperature $T=W/140$ ($W=$ bandwidth). Calculating with 
significantly higher precision than in previous work, we show that the 
hysteresis below $T_{\mbox{\tiny IPT}}\simeq 0.022W$, reported in earlier 
studies, disappears. Hence the transition is found to be {\em continuous\/} 
rather than discontinuous down to at least $T=0.325T_{\mbox{\tiny IPT}}$. We 
also study the changes in the density of states across the transition,
which illustrate that the Fermi liquid breaks down before the gap opens.
}}

\maketitle

\narrowtext

The explanation of the nature of the Mott-Hubbard metal-insulator
transition, i.e., the transition between a paramagnetic metal and a
paramagnetic insulator, is one of the classic and fundamental problems in 
condensed matter physics~\cite{Mott,Gebhard}. Metal-insulator transitions 
of this type  are, for example, found in transition metal oxides with
partially filled bands near the Fermi level. For such systems band theory
typically predicts metallic behavior. The most famous example is 
V$_{2}$O$_{3}$ doped with Cr~\cite{McWhR,McWhetal,RMcWh}. In particular, 
in (V$_{0.96}$Cr$_{0.04}$)$_{2}$O$_{3}$ the metal-insulator transition is 
{\em first-order\/} below $T_{c}\simeq 380$K \cite{McWhetal}, with 
discontinuities in the ratio of the lattice parameters (the two phases 
being isostructural, however) and in the conductivity, accompanied by 
hysteresis. 

The Mott-Hubbard transition is caused by electron-electron repulsion. 
The fundamental features of this transition are traditionally expected 
\cite{RMcWh,Mott} to be explainable in terms of the half-filled
single-band Hubbard model~\cite{HubbardI,Gutzwiller,Kanamori},
\[
H=-t\sum_{{\bf (ij)},\sigma} 
 c_{{\bf i}\sigma}^{\dagger}c_{{\bf j}\sigma}^{\phantom{\dagger}}+
         U\sum_{{\bf i}} n_{{\bf i}\uparrow}^{\phantom{\dagger}} 
         n_{{\bf i}\downarrow}^{\phantom{\dagger}}\; ,
\]
which describes electrons hopping on a lattice, interacting with each other 
through on-site Coulomb repulsion. 

On the basis of this model the Mott-Hubbard transition was studied
intensively over the last 35 years. Important early results were
obtained by Hubbard~\cite{Hub3} with a Green function decoupling scheme,
and by Brinkman and Rice~\cite{BR} with the Gutzwiller variational 
method~\cite{Gutzwiller}, both at $T=0$. Hubbard's approach yields a 
continuous splitting of the band into a lower and upper Hubbard band, but 
does not describe quasiparticle features. By contrast, the 
Gutzwiller-Brinkman-Rice approach concentrates on the low-energy behavior, 
the transition being monitored by the disappearance of the quasiparticle peak,
but does not produce the upper and lower Hubbard bands. A unified approach, 
treating both the low energy and high energy features on equal footing, has 
recently become possible within the Dynamical Mean-Field Theory (DMFT) 
\cite{GKreview}, which provides the exact solution of the Hubbard model in 
the limit of infinite dimensionality (or coordination number) \cite{MV}. 
The complicated structure of the self-consistent DMFT-equations makes an 
analytic solution untractable and hence one has to resort to approximate 
techniques. In the last few years Georges, Kotliar and collaborators 
performed detailed investigations of the metal-insulator transition scenario 
within the DMFT, by employing iterated perturbation theory (IPT), exact 
diagonalization (ED) of small systems, quantum Monte-Carlo (QMC) simulations
and, at $T=0$, a projective self-consistent technique (PSCT) 
\cite{GKreview,Moeller95}. While the overall transition scenario reported by 
these authors indeed combines essential features of the early approaches,
they find the transition to be {\em discontinuous\/}, with hysteresis, due 
to a coexistence regime between the metallic and the insulating phase. For 
all finite temperatures $T<T_{\mbox{\scriptsize IPT}}$ the quasiparticle 
weight disappears abruptly and the gap between the Hubbard bands opens 
discontinuously as a function of $U$. Hence these authors argued that the 
experimentally observed metal-insulator transition in V$_{2}$O$_{3}$ can 
already be understood using a purely electronic correlation model. 

At $T=0$, numerical renormalization group (NRG) studies \cite{RBTP} 
also found hysteresis and a value for the critical interaction of $U_c=5.86$ 
(in our units, see below), which agrees with the results of the PSCT 
\cite{Moeller95}. Nevertheless, the existence of a preformed gap at $T=0$ and 
the corresponding separation of energy scales on which the PSCT is based were 
recently disputed by one of us \cite{Kehrein98}. Finally, a {\em continuous\/} 
transition with a considerably lower $U_c$ was recently observed within the 
random dispersion approximation (RDA \cite{Gebhard,RNFG}). Clearly the 
Mott-Hubbard transition scenario is still very controversial.

It is the purpose of this paper to carefully re-examine the nature of the 
metal-insulator transition within DMFT at finite temperatures. The examination 
of a transition region requires a technique with sufficient precision. Since 
IPT is a rather {\em ad hoc\/} approximation scheme, its qualitative and 
quantitative accuracy is uncertain. On the other hand, for $T>0$ ED is limited 
to quite small systems ($\leq 7$ sites), so that finite-size effects may be 
considerable. Indeed, although both techniques predict the metal-insulator 
transition to be discontinuous, their {\em quantitative\/}  predictions differ
substantially (see below). In fact, their respective regions of hysteresis do 
not even overlap much. To resolve these discrepancies we perform 
finite-temperature QMC calculations, using two different codes to reduce 
possible systematic errors. Although QMC is limited to not too low $T$- 
and not too large $U$-values, it is still the best understood and most 
thoroughly tested technique presently available for the solution of the 
DMFT-equations. For comparison with Ref.\ \cite{GKreview} we focus on the 
Hubbard model with a semi-elliptical non-interacting density of states (DOS): 
$N(E)=\sqrt{4-(E/t^{*})^{2}}/(2\pi t^{*})$ if $|E|\leq 2t^{*}$ and zero 
elsewhere. This DOS is realized, e.g., on a Bethe lattice with hopping 
amplitude scaled as $t=t^{*}/\sqrt{Z}$, where $Z\rightarrow\infty$ is the 
coordination number. In the following we set $t^{*}=1$. In order to study the 
Mott-Hubbard transition, we restrict our calculations to the paramagnetic 
phase and exclude symmetry breaking. The solution of the Hubbard model is then
determined by the following single-site effective action 
\cite{GKreview,BraMie,Janis,MJPRL,Georges92}:
\begin{eqnarray*}
S_{\rm eff}&=&-\int_0^\beta d\tau\, d\tau'\, \sum_\sigma
c^\dagger_{\sigma}(\tau) {\cal G}^{-1}_0(\tau-\tau') 
c^{\phantom{\dagger}}_{\sigma}(\tau') \\
&+&U\int_0^\beta d\tau 
\left( c^\dagger_\uparrow(\tau) c^{\phantom{\dagger}}_\uparrow(\tau)
-\frac{1}{2}\right) 
\left( c^\dagger_\downarrow(\tau) c^{\phantom{\dagger}}_\downarrow(\tau)
-\frac{1}{2}\right) \; . 
\end{eqnarray*}
In particular, the on-site Green function $G_{\bf ii}(\tau-\tau')$ 
of the Hubbard model is identical to the single-site Green function 
$G(\tau-\tau')\equiv -\langle {\cal T}c^{\phantom{\dagger}}(\tau)
c^\dagger(\tau')\rangle_{S_{\rm eff}}$, which is implicitly determined
by $S_{\rm eff}$ and the self-consistency relation 
${\cal G}_{0}(i\omega_n)=\left[ i\omega_n+\mu-G(i\omega_n)\right]^{-1}$,
where $\omega_{n}=(2n+1)\pi T$. The key point in DMFT is thus the accurate 
calculation of the single-site Green function $G(\tau)$. For this purpose we 
use QMC-simulations, which are essentially exact though computationally 
expensive \cite{MJPRL,UJV}. After discretizing the imaginary time into 
$\Lambda$ time slices of length $\Delta\tau =\beta /\Lambda$ and performing a 
Hubbard-Stratonovich transformation which introduces auxiliary Ising spins, 
the DMFT-equations are solved by iteration. The number of proposed flips of 
Ising spins (``sweeps'') per iteration will be important here. Each iteration 
has as input the ``old'' self-energy $\Sigma_{\rm old}(i\omega )$ and as 
output a ``new'' self-energy $\Sigma_{\rm new}(i\omega )$. The rate of change 
in the iteration procedure is measured by $\eta\equiv\Lambda^{-1}\sum_{n}
|\Sigma_{\rm old}(i\omega_{n})-\Sigma_{\rm new}(i\omega_{n})|$.
Experience shows that, for most purposes (e.g., calculation of thermodynamic
quantities outside the critical regime), convergence is reached if
$\eta\leq 10^{-3}$. Physical properties are finally obtained by extrapolation 
of the simulation results for various $\Delta\tau$-values to the limit 
$\Delta\tau\to 0$.

\begin{figure}[t]
\begin{center}
\leavevmode
\epsfxsize=7cm
\epsfysize=6cm
\epsfbox{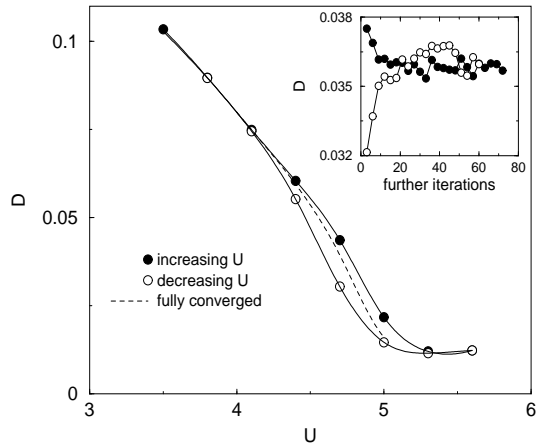}
\caption{QMC-results for the double occupancy $D$ at $T=1/20$
using the criterion $\eta\leq 10^{-3}$. The error bars are smaller than the 
data points. All lines are guides to the eye only. Inset: Vanishing of the 
hysteresis as a function of the number of further iterations at $T=0.05$, 
$U=5$ and $\Delta\tau =0.3$.}
\end{center}
\end{figure}

\begin{figure}[t]
\begin{center}
\leavevmode
\epsfxsize=7cm
\epsfysize=6cm
\epsfbox{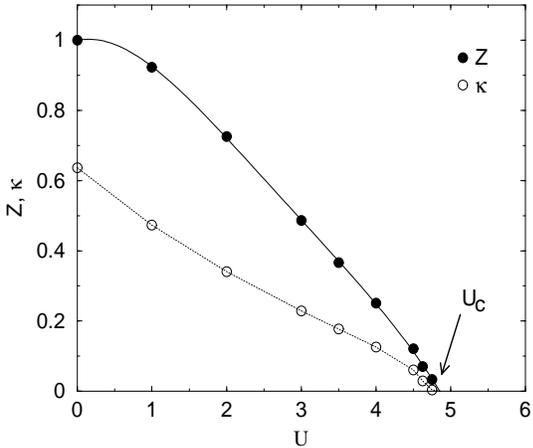}
\caption{Compressibility $\kappa$ and quasiparticle renormalization factor 
$Z=m/m^{*}$ vs.\ $U$ for $T=1/15$. All lines are guides to the eye only.}
\end{center}
\end{figure}

We found that straightforward implementation of the QMC algorithm with
$\eta\leq 10^{-3}$ indeed leads to the apparent convergence of {\em two\/}
solutions, i.e., to hysteresis. As an example we present results for the 
double occupancy $D=\langle n_{{\bf i}\uparrow} n_{{\bf i}\downarrow}\rangle$ 
in Fig.\ 1. The upper curve (solid dots) shows results for increasing coupling 
strength, where we use the self-energy calculated for interaction $U$ as input 
for the calculation for $U+\Delta U$ (here $\Delta U=0.3$). Similarly, the 
lower curve (open dots) shows results for decreasing $U$. At each $U$-value 
the iteration procedure was terminated as soon as the condition 
$\eta\leq 10^{-3}$ was satisfied, which typically happens after only a few 
iterations. In this way we are able to {\em qualitatively reproduce the 
hysteresis found in Ref.\/} \cite{GKreview} by means of a standard 
QMC-algorithm with a standard convergence criterion. However, as will be shown 
below, in the case of the Mott-Hubbard transition the criterion 
$\eta\leq 10^{-3}$ is not sufficient to guarantee convergence to the true 
solution.

For this we investigate the stability of the two solutions independently
under further iterations. As an example we consider the solution at
$T=1/20$ and $U=5$. The data for $\Delta\tau =0.3$ are presented in the
inset of Fig.\ 1. In order to obtain maximum accuracy we use a large number
of sweeps per iteration ($2-3\cdot 10^{5}$). We find that ($i$) {\em both\/} 
solutions at $U=5$ in Fig.\ 1 are unstable, ($ii$) a new stable solution is 
reached after approximately 20 more iterations, and ($iii$) the solution
with the larger number of sweeps and time slices (solid dots) reaches 
equilibrium sooner and has smaller fluctuations. During the iteration process 
$\eta$ fluctuates around $\langle\eta\rangle \simeq 4\cdot 10^{-4}$ 
and gives essentially no information about the distance from equilibrium. We 
obtained similar results for a dense grid of other values for $U$ and 
$\Delta\tau$. Extrapolating to $\Delta\tau\to 0$ and combining the results for
various $U$, we obtain a smooth curve $D(U)$ {\em without\/} hysteresis 
(dashed curve in Fig.\ 1). The same happens at higher temperatures 
($T=1/10$, $1/13$ and $1/15$) and also at lower temperatures, such as 
$T=1/30$ and $T=1/35$.

We also studied the quasiparticle renormalization factor $Z=m/m^{*}$ and the 
compressibility $\kappa$ at various temperatures. After 15-25 further 
iterations, none of these quantities shows hysteresis any more. The results 
for $T=1/15$ are shown in Fig.\ 2. We locate the Mott-Hubbard transition
at the interaction strength $U_{c}$ where $Z(U)$ and $\kappa (U)$ essentially
vanish. The resulting phase diagram is plotted in Fig.\ 3, where the 
corresponding IPT- and ED-results\cite{GKreview,Hofstetter} and the 
$U_{c}$-values at $T=0$ obtained by PSCT \cite{Moeller95}, NRG 
\cite{RBTP} and RDA \cite{RNFG} are also shown. There is a clear quantitative 
and qualitative discrepancy between our numerically exact QMC-data and both 
the ED- and the IPT-results. According to IPT, the transition below 
$T=T_{\mbox{\scriptsize IPT}}\simeq 0.088$ is discontinuous. By contrast, we 
find that the transition from the metal to the insulator is {\em continuous\/}
down to at least $T=1/35\simeq 0.325\, T_{\mbox{\scriptsize IPT}}$. 

\begin{figure}[t]
\begin{center}
\leavevmode
\epsfxsize=8.5cm
\epsfysize=6cm
\epsfbox{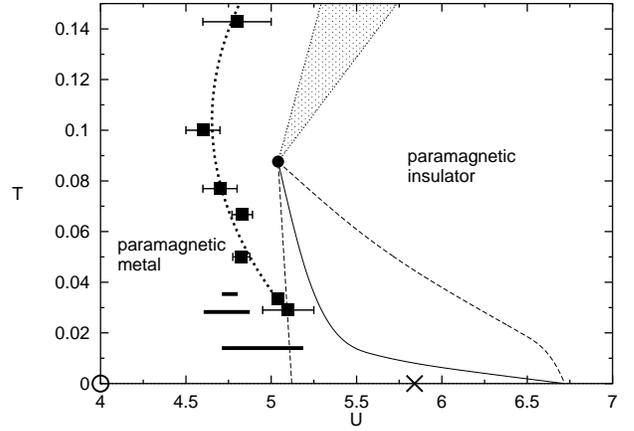}
\caption{Phase diagram of the Hubbard model (paramagnetic phase only). Solid 
squares: continuous metal-insulator transition calculated with QMC (error bars 
include both the statistical errors and the uncertainty in $U_{c}$ due to 
finite temperatures). The dotted line is a guide to the eye only. Broad 
horizontal lines: coexistence region within ED 
\protect\cite{GKreview,Hofstetter}. Dashed lines: coexistence region 
within IPT; the line of first-order transitions (full curve) ends at 
$T_{\mbox{\scriptsize IPT}}$ (solid circle) \protect\cite{GKreview}. The 
shaded area is a crossover region. Also shown are the $U_{c}$-values from 
PSCT/NRG ({\sf X}) and RDA ({\sf O}).}
\end{center}
\end{figure}

\begin{figure}[t]
\begin{center}
\leavevmode
\epsfxsize=7cm
\epsfysize=6cm
\epsfbox{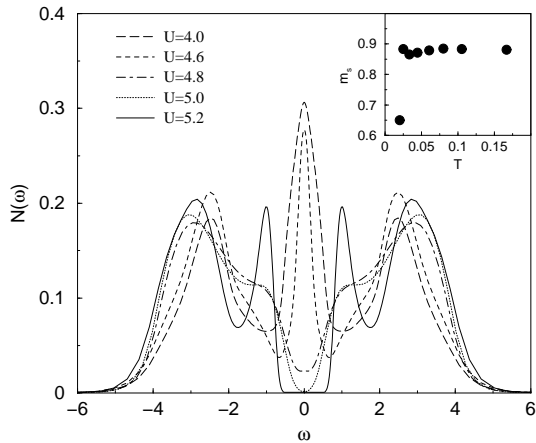}
\caption{QMC/MEM-results for the DOS (Bethe lattice) for various values of $U$ 
at $T=1/20$, and $\Delta\tau=0.017$. The gap opens at $U_{g}\simeq 5.0$.  The 
temperature-dependence of the screened local moment $m_{s}$ (the product of 
the temperature and the local susceptibility) is shown in the inset for 
$U=4.8$.}
\end{center}
\end{figure}

More information concerning the Mott-Hubbard transition can be obtained from
the DOS, which we calculated from the converged data for the Green function
using the Maximum Entropy Method (MEM\cite{MEM}). Our results for the DOS as
a function of $U$ at $T=1/20$ are presented in Fig.\ 4. Upon increase of $U$, 
the DOS develops a well-defined central peak and shoulders, with the peak
pinned at its Fermi liquid value. The peak rapidly collapses at
$U\simeq U_{c}$ (here: $U_{c}\simeq 4.8$), leaving an incomplete gap
\cite{FLbreakdown}. Beyond $U_{c}$ the system remains semi-metallic until at
$U_{g}>U_{c}$ (here: $U_{g}\simeq 5.0$) an actual gap opens (within numerical 
accuracy). In order to better understand the region $U_{c}<U<U_{g}$ we 
studied the temperature dependence of the screened local moment (inset to 
Fig.\ 4) and of the spectrum (not shown) at $U=4.8$, starting from $T=1/20$. 
Upon increase of $T$ the incomplete gap and the screened local moment remain 
essentially unchanged. Upon decrease of $T$ the central peak and the 
Fermi-liquid behavior are rapidly restored, and the screened moment falls
dramatically. The incomplete gap and the temperature-independence of the
screened moment imply that for $U_{c}<U<U_{g}$ there are few electronic
states at the Fermi energy to screen the spins. The behavior in this region 
can be interpreted as emanating from a quantum critical point at $T=0$,
characterized by a vanishing DOS at the Fermi level, $N(0)=0$. The depletion 
of screening states obtains a natural interpretation in terms of Nozi\`{e}res'
``exhaustion'' scenario \cite{Nozieres}, which was recently found to be
realized in the periodic Anderson model \cite{Nikietal}.

To clarify the influence of band-structure effects we also performed 
calculations for a {\em hypercubic lattice\/} in $d=\infty$ including 
next-nearest-neighbor hopping amplitudes 
$t^{\prime }=t^{\prime }{}^{\ast }/\sqrt{2d(d-1)}$ 
\cite{JSthesis,MuellerHartmann89a}. Choosing 
$t^{\prime }{}^{\ast }/t^{\ast }<0$ in order to obtain a finite lower band 
edge, we find the hysteresis effects to be strongly suppressed by frustration,
e.g., no initial hysteresis was observed for 
$t^{\prime }{}^{\ast }/t^{\ast }\alt-0.25$. The phase diagram for 
$t^{\prime }=0$, previously obtained \cite{Jarrell_Pruschke} using QMC and
perturbation theory in $t^{\ast}$ (NCA), is qualitatively similar to that
for the Bethe lattice in Fig.\ 3.

In summary, we demonstrated that, for temperatures down to 
$T_{\mbox{\scriptsize min}}=1/35\simeq 0.325\, T_{\mbox{\scriptsize IPT}}$, 
the coexistence region characteristic of a first-order metal-insulator 
transition disappears in sufficiently careful QMC simulations. Further 
conclusions about the order of the transition cannot safely be made. 
Physically we would expect the transition to be a smooth (broadened by finite 
temperatures) but rapid {\em crossover\/} between $U_{c}$ (where the Fermi
liquid breaks down) and $U_{g}>U_{c}$ (where the gap opens). We cannot rule 
out the existence of a first-order transition at even lower temperatures and 
note that our results may be smoothly connected to those from PSCT 
\cite{Moeller95} and NRG \cite{RBTP} at $T=0$. However, we found no evidence 
for the existence of a preformed gap at low temperatures, in line with the 
transition scenarios of Refs.\ \cite{Gebhard,Kehrein98,Nozieres}. 
-- For a bandwidth of 0.8 eV \cite{estimate} our results imply that down to 
$T_{\mbox{\scriptsize min}}\simeq 70$K the metal-insulator transition is 
{\em continuous\/}. Since in the experiment on Cr-doped V$_{2}$O$_{3}$ the 
transition is {\em first-order\/} below $T_{c}\simeq 380$K \cite{McWhetal}, 
an explanation of the nature of this transition apparently requires the 
inclusion of other degrees of freedom, such as lattice \cite{Majumdar}, 
orbital \cite{Castellani,Bao} and possibly higher-spin \cite{Ezhov} effects.

We are grateful to R. Bulla, F. Gebhard, K. Held, W. Hofstetter, S. Horn, 
R. Noack, G. Kotliar, H.-R. Krishnamurthy, W. Krauth, M. Rozenberg, M. Ulmke, 
and F. C. Zhang for discussions. Support by NSF grants DMR--9704021 and 
DMR--9357199 and the Ohio Supercomputer Center (MJ) and by the John von 
Neumann Institute for Computing, J\"{u}lich, (JS, NB, DV) is acknowledged.


\begin{references}
\bibitem{FLbreakdown} The breakdown of the Fermi liquid does not coincide
                      with the transition to the insulator; similar behavior
                      is found for normal liquid $^{3}$He upon increasing
                      pressure.
\bibitem{estimate} This value was argued in Ref.\ \cite{GKreview}, p.\ 81, to
                   be close to the experimental bandwidth.
\end{references}
\end{document}